\documentclass[aps,prd,preprint,tightenlines,groupedaddress,showpacs]{revtex4}
\usepackage{epsf,epsfig,graphicx,color,hyperref}
\newcommand{\alpr}{\alpha}

\newcommand{\vk}{\vec{k}}
\newcommand{\vl}{\vec{l}}

\newcommand{\vkdotvl}{\vec{k} \cdot \vec{l}}

\begin{document}

\title{Supernova Bounds on Weinberg's Goldstone Bosons}

\author{Wai-Yee Keung}
\email{keung@uic.edu}
\affiliation{Physics Department, University of Illinois at Chicago, Chicago IL 60607, USA}
\author{Kin-Wang Ng}
\email{nkw@phys.sinica.edu.tw}
\affiliation{Institute of Physics, Academia Sinica, Taipei, Taiwan 11529, R.O.C.}
\author{Huitzu Tu}
\email{huitzu@phys.sinica.edu.tw}
\affiliation{Institute of Physics, Academia Sinica, Taipei, Taiwan 11529, R.O.C.}
\author{Tzu-Chiang Yuan}
\email{tcyuan@phys.sinica.edu.tw}
\affiliation{Institute of Physics, Academia Sinica, Taipei, Taiwan 11529, R.O.C.}

\begin{abstract}
Recently, Weinberg proposed a scenario where Goldstone bosons may be masquerading 
as fractional cosmic neutrinos. We calculate the energy loss rates through the emission of
these Goldstone bosons in a post-collapse supernova core.
Invoking the well established emissivity bound from the Supernova 1987A observations
and simulations, 
we find that nuclear bremsstrahlung processes can notably impose a 
bound on the Goldstone boson coupling to the Standard Model Higgs, $g$, 
dependent on the mass of the associated radial field, $m_r$.
For $m_r$ large enough compared with the temperature in the post-collapse 
supernova core, our bound is $|g| \lesssim 0.011\, (m_r / 500~{\rm MeV})^2$, very competitive to that derived from collider experiments.
\end{abstract}


\date{\today}

\maketitle

The cosmic microwave background (CMB) radiation, if combined with other observational data, 
can be used to constrain the effective number of light neutrino species.
The WMAP9 data combined with eCMB, BAO, and $H_0$ measurements has inferred 
$N_\nu = 3.55^{+0.49}_{-0.48}$ at 68$\%$ CL~\cite{Hinshaw:2012aka}.
Latest Planck data combined with WP, highL, BAO, and $H_0$ measurements gives 
$N_\nu = 3.52^{+0.48}_{-0.45}$ at 95$\%$ CL~\cite{Ade:2013zuv}.
These bounds are consistent with that from the big bang nucleosynthesis (BBN) 
$N_\nu = 3.71^{+0.47}_{-0.45}$ (see e.g. Ref.~\cite{Steigman:2012ve}).
On the other hand, the standard scenario with three active, massless neutrinos
predicts $N_\nu = 3.046$ at the CMB epoch~\cite{Mangano:2001iu}.

The latest CMB data is in excellent agreement with the standard value and leave less and less
room for extra weakly-interacting light species.
Recently, Weinberg ~\cite{Weinberg:2013kea} has investigated whether Goldstone bosons 
can still be masquerading as fractional cosmic neutrinos.
The motivation is that they would be massless or nearly massless, and
their characteristic derivative coupling would make them very weakly-interacting
at sufficiently low temperatures.
The most crucial criterion is that those Goldstone bosons have to decouple from the thermal
bath early enough so that their temperature is lower than that of the neutrinos.
A simple extended Higgs sector in the Standard Model (SM) has been proposed to realize
this idea such that the Goldstone bosons contribute significantly to the effective number of light species.
The thermal history of these Goldstone bosons depends crucially on their coupling to the Standard
Model Higgs field and and the mass of the radial field.
An upper bound on the coupling constant can be quickly derived using the limit on the invisible decay width of the SM Higgs.
In this work we will examine the viability of this scenario by considering the cooling
of a post-collapse supernova core, such as the Supernova 1987A.

Let us first briefly summarize Weinberg's model~\cite{Weinberg:2013kea} 
following the convention of Ref.~\cite{Cheung:2013oya}.
Consider the simplest possible broken continuous symmetry, a global $U (1)$ symmetry associated with
the conservation of some quantum number $W$.
A single complex scalar field $S(x)$ is introduced for breaking this symmetry spontaneously.
With this field added to the SM, the Lagrangian is
\begin{eqnarray}
\label{Lagrangian}
  {\mathcal L}= (\partial_\mu S^\dagger ) (\partial^\mu S ) + \mu^2 S^\dagger S 
  - \lambda (S^\dagger S)^2  - g (S^\dagger S) (\Phi^\dagger \Phi) 
  + {\cal L}_{\rm SM}\, ,  
\end{eqnarray}
where $\Phi$ is the SM Higgs doublet, $\mu^2$, $g$, and $\lambda$ are real constants, 
and $\mathcal{L}_{\rm SM}$ is the usual SM Lagrangian. 
One separates a massless Goldstone boson field $\alpha (x)$ and a massive radial field $r (x)$ 
in $S(x)$ by defining
\begin{equation}
  S(x) = \frac{1}{\sqrt 2}\left( \langle r \rangle + r (x) \right) e^{2 \imath \alpha (x)}\, ,
\end{equation}
where the fields $r (x)$ and $\alpha (x)$ are real. 
In the unitary gauge, one sets $\Phi^{\rm T} = (0 \, , \, \langle\varphi \rangle + \varphi (x) )/\sqrt 2$
where $\varphi (x)$ is the physical Higgs field.
The Lagrangian in Eq.~(\ref{Lagrangian}) thus becomes
\begin{eqnarray}
\label{Lagrangian2}
{\mathcal L} & = & \frac{1}{2} (\partial_\mu r )(\partial^\mu r) 
  + \frac{1}{2} \frac{ (\langle r \rangle + r)^2}{\langle r \rangle^2}
   (\partial_\mu \alpha )(\partial^\mu \alpha) 
  + \frac{\mu^2}{2} ( \langle r \rangle + r )^2 
  - \frac{\lambda}{4} ( \langle r \rangle + r )^4 \nonumber \\
  && -  \, \frac{g}{4} ( \langle r \rangle + r )^2 ( \langle \varphi \rangle + \varphi )^2 
  + \mathcal{L}_{\rm SM} \, .
\end{eqnarray}
In Eq.~(\ref{Lagrangian2}), we have replaced 
$\alpha(x) \to \alpha(x)/(2\langle r \rangle )$ in order to achieve a canonical
kinetic term for the $\alpha(x)$ field.
In this model, the interaction of the Goldstone bosons with the SM particles arises entirely
from a mixing of the radial boson with the Higgs boson via the following mixing angle
\begin{equation}
   \tan 2 \theta = \frac{2 g \left< \varphi \right> \left< r \right>}{m^2_\varphi - m^2_r}\, .
\end{equation}
The $\varphi$-$r$ mixing allows the SM Higgs boson to decay into a pair of the Goldstone bosons 
with the decay width 
\begin{equation}
   \Gamma_{\varphi \rightarrow 2 \alpha} = \frac{g^2 \left<\varphi \right>^2 m^3_\varphi}
   {32 \pi\, (m^2_\varphi - m^2_r)^2}\, .
\label{decaywidth}
\end{equation}
For $\left<\varphi \right> = 247$ GeV, $m_\varphi = 125$ GeV, and assuming $m_r \ll m_\varphi$,
one obtains a constraint of $|g| \lesssim 0.018$. 
In Ref.~\cite{Cheung:2013oya} it is pointed out that by including the $\varphi \rightarrow r r$ channel, the constraint can be improved to $|g| \lesssim 0.011$.
Further collider signatures of this model have been investigated therein and in
Ref.~\cite{Anchordoqui:2013bfa}.

From the mixing term $- g \left<\varphi \right> \left<r \right> \varphi r$ 
and the interaction term $(1/ \left<r \right>)\, r\, \partial_\mu \alpha\, \partial^\mu \alpha$ 
in the Lagrangian (Eq.~(\ref{Lagrangian2})) as well as the SM Higgs-fermion coupling 
$-m_f \varphi \bar f f /\langle \varphi \rangle$,
an effective interaction between the Goldstone bosons and any SM fermion $f$,
\begin{equation}
   +g\, m_f\, \bar{f} f\, \varphi r\, \partial_\mu \alpha\, \partial^\mu \alpha\, ,
\label{ealpha}
\end{equation}
is produced. 
In the early universe, the Goldstone bosons remain in thermal equilibrium
via the processes $\alpr \alpr \leftrightarrow \bar{f} f$, where $f$ are SM fermions in the 
thermal bath.
If the Goldstone bosons freeze out before the muon annihilation occurs, they contribute about 0.39 to the effective number of neutrino types in the era before recombination.
Weinberg has made a crude estimate which shows that for $g =0.005$ the Goldstone bosons decouples 
at muon annihilation for $m_r \approx 500~{\rm MeV}$.
While a more accurate calculation is underway~\cite{longpaper}, in this work
we will use $m_r=500~{\rm MeV}$ as a benchmark.

Now we turn to supernova cooling.  
The observed duration of neutrino burst events from Supernova 1987A in several detectors 
confirmed the standard picture of neutrino cooling of post-collapse supernova.
In the second phase of neutrino emission, a light particle which interact even more weakly than
neutrinos could lead to more efficient energy loss and shorten the neutrino burst duration.
Demanding that the novel cooling agent should not have affected the  
total cooling time significantly,
an upper bound on their emissivity can be derived~\cite{Raffelt:1990yz,Raffelt:2006cw}
\begin{equation}
\label{emissivity_bound}
   \epsilon_X \equiv \frac{Q_X}{\rho} \lesssim 10^{19}\, {\rm erg} \cdot {\rm g}^{-1} \cdot
  {\rm s}^{-1} = 7.324 \cdot 10^{-27}\, {\rm GeV}\, .
\end{equation}
It has been used exhaustively in the literature to constrain the properties of exotic particles,
notably the axions~\cite{Raffelt:1987yt,Turner:1987by,Mayle:1987as}, 
right-handed neutrinos~\cite{Raffelt:1987yt}, 
Kaluza-Klein gravitons~\cite{Hannestad:2001xi}, and unparticles~\cite{Freitas:2007ip} etc.
To apply this bound one assumes typical core conditions, i.e. 
a density $\rho \approx 3 \cdot 10^{14}$ g/cm$^3$ and 
a temperature $T \sim 30$ MeV.
Denote the number densities of neutron, proton, electron, and electron neutrino, by
$n_n$, $n_p$, $n_e$, and $n_{\nu_e}$, respectively.
From the baryon density ($n_B = n_n + n_p$),
charge neutrality ($n_p = n_e$), and $\beta$-equilibrium ($\mu_p + \mu_e = \mu_n + \mu_{\nu_e}$),
one can find the chemical potential ($\mu_i$) of each particle for a fixed lepton fraction
($Y_L = (n_e + n_{\nu_e})/n_B$.)
For $T =30$ MeV and $Y_L=0.3$, they are $\mu_n=971$~MeV, $\mu_p=923$~MeV, $\mu_e=200$~MeV, 
and $\mu_{\nu_e}=152$~MeV, respectively. 
The degeneracy parameter for the neutron is $\eta_n \equiv (\mu_n - m_n)/T \approx 1.05$ in this case, corresponding to neither strongly non-degenerate nor degenerate case.
On the other hand, the electrons are highly degenerate.

Stellar energy loss due to Goldstone boson pair 
emission had been considered for the Compton-like process \cite{Chang:1988fz}.
Here, from their effective interaction with the SM fermions (Eq.~(\ref{ealpha})),
the Goldstone bosons can be produced in electron-positron pair annihilation  
$e^+ e^- \rightarrow \alpha \alpha$, in photon scattering $\gamma \gamma \rightarrow \alpha \alpha$
and in nuclear bremsstrahlung processes $N N\rightarrow N N \alpha \alpha$.
For the $e^+ (p_1)\, e^- (p_2) \leftrightarrow \alpha (q_1)\, \alpha (q_2)$ process, the amplitude squared is
\begin{equation}
\label{eq:Msq_mumu}
  \frac{1}{4}\, \sum_{\rm spins}\, |\mathcal{M}_{e^+ e^- \rightarrow \alpha \alpha}|^2 = 
  \frac{g^2\, m^2_e}{(s-m^2_\varphi)^2\, (s-m^2_r)^2}\, (q_1 \cdot q_2)^2\, 
  \cdot 4 \left[(p_1 \cdot p_2) - m^2_e \right]\, ,
\end{equation}
where $s= (p_1 + p_2)^2 = (q_1 + q_2)^2$ is the center-of-mass energy squared.
The energy loss rate due to this process is 
\begin{eqnarray}
   Q_{e^+ e^- \rightarrow \alpr \alpr} &=& \frac{1}{2!} 
   \int \frac{d^3\vec{q_1}}{(2\pi)^3 2 \omega_1} 
   \frac{d^3 \vec{q_2}}{(2\pi)^3 2 \omega_2}\, 
   \int \prod^{2}_{i=1} \frac{2\, d^3 \vec{p_i}}{2 E_i\, (2 \pi)^3}\,
   \frac{1}{4}\, \sum_{\rm spins}\,
   |\mathcal{M}_{e^+ e^- \rightarrow \alpr \alpr}|^2 \nonumber \\
   && \times (2\pi)^4 \delta^4 (p_1 + p_2 - q_1 - q_2)\, f_1 f_2\, (\omega_1 + \omega_2)\, ,  
\end{eqnarray}
where $f_1 (\vec{p}_1) = (e^{(E_1 + \mu_e) / T} + 1)^{-1}$ and 
$f_2 (\vec{p}_2) = (e^{(E_2 - \mu_e) / T} + 1)^{-1}$
are the distribution functions for the positron and the electron, respectively.
A symmetry factor of $1/2!$ is included for the identical particles in the final state.
In the large $m_r$ limiting case where $m^2_r \gg s$, the integrand can be 
expanded as 
\begin{equation}
\label{mr_large}
   \frac{(q_1 \cdot q_2)^2}{(2 q_1 \cdot q_2 - m^2_r)^2} \approx 
   \frac{(q_1 \cdot q_2)^2}{m^4_r}\, \left(1 - 4 \frac{(q_1 \cdot q_2)}{m^2_r} + ... \right)\, .
\end{equation}
Taking only the leading term and performing the $d^3 \vec{q}_1 d^3 \vec{q}_2$ integral 
analytically, we obtain 
\begin{equation}
   \int \frac{d^3 \vec{q}_1}{\omega_1} \frac{d^3 \vec{q}_2}{\omega_2}\, 
   \frac{(q_1 \cdot q_2)^2}{m^4_r}\,
   \delta^4 (p_1 + p_2 - q_1 - q_2) = \frac{\pi}{2} \frac{(p_1 + p_2)^4}{m^4_r}\, ,
\end{equation}
analogous to the Lenard's Identity for the $e^+ e^- \rightarrow \nu \bar{\nu}$ process~\cite{Lenard:1953}.
Then following Ref.~\cite{Yakovlev:2000jp}, we define these two dimensionless functions
\begin{equation}
   U_k \equiv \frac{1}{\pi^2}\, \int^\infty_0 \frac{|\vec{p}_1|^2 d |\vec{p}_1|}{T^3}\,
   \left(\frac{E_1}{T} \right)^k\, f_1 (\vec{p}_1)\, , \hspace{0.5cm}
   \Phi_k \equiv \frac{1}{\pi^2}\, \int^\infty_0 \frac{|\vec{p}_2|^2 d |\vec{p}_2|}{T^3}\,
   \left(\frac{E_2}{T} \right)^k\, f_2 (\vec{p}_2)\, ,
\end{equation}
so that the energy loss rate can be expressed as
\begin{eqnarray}
   Q_{e^+ e^- \rightarrow \alpr \alpr} &=& \frac{T^{11}}{16 \pi} 
   \left(\frac{g\, m_e}{m^2_\varphi\, m^2_r} \right)^2\, \Big\{2 (U_2\, \Phi_3 +  U_3\, \Phi_2)
   + \frac{1}{3} (U_1\, \Phi_2 + U_2\, \Phi_1) - (U_0\, \Phi_3 + U_3\, \Phi_0) \nonumber \\
   && - \frac{1}{3} (U_0\, \Phi_1 + U_1\, \Phi_0) - 
   \frac{1}{3} (U_{-1}\, \Phi_2 + U_2\, \Phi_{-1}) -
   \frac{2}{3} (U_{-1}\, \Phi_0 + U_0\, \Phi_{-1}) \Big\}\, .
\end{eqnarray}
Evaluating the $U_k$, $\Phi_k$ functions numerically for the typical supernova core condition
$\rho = 3 \cdot 10^{14}~{\rm g}/{\rm cm}^3 $, $T=30~{\rm MeV}$ and $\mu_e = 200~{\rm MeV}$, 
we find the emissivity due to the process $e^+ e^- \rightarrow \alpr \alpr$ is
\begin{equation}
   \epsilon_{e^+ e^- \rightarrow \alpr \alpr} = \frac{Q_{e^+ e^- \rightarrow \alpr \alpr}}{\rho}
   = 1.73 \cdot 10^{-28}~{\rm GeV}\, g^2\, \left(\frac{m_r}{500~{\rm MeV}} \right)^{-4}\, .
\end{equation}
One sees that for $m_r$ around $500~{\rm MeV}$, 
even with $g \approx 0.018$ saturating the collider bound, contribution from Goldstone boson emission to supernova cooling is far from competing with that from neutrino emission.

The energy loss rate for the photon scattering process can be calculated similarly.
The amplitude squared for the process $\gamma (p_1)\, \gamma (p_2) \rightarrow 
\alpr (q_1)\, \alpr (q_2)$ is
\begin{equation}
   |\mathcal{M}_{\gamma \gamma \rightarrow \alpr \alpr}|^2 = 
   \left(\frac{\alpha}{4 \pi} \right)^2\, \frac{16\, G_F}{\sqrt{2}}\, |F|^2\, (q_1 \cdot q_2)^2\,
   \frac{g^2 \left<\varphi \right>^2}{(s-m^2_\varphi)^2}\, 
   \frac{(p_1 \cdot p_2)^2}{(s-m^2_r)^2}\, ,
\end{equation}
where $\alpha$ is the fine-structure constant, and $G_F$ is the Fermi constant.
The form factor $F$ enters through the amplitude for the SM Higgs decay to two photons (see e.g. Ref.~\cite{Marciano:2011gm}).
It is a function of the center-of-mass energy squared $s$.
At the energies attainable at the typical temperature in the post-collapse supernova core, 
$F$ receives contribution from the $W$ boson, charged leptons and hadrons, 
$F = F_W + \sum_l F_l + F_{\rm had}$~\cite{Ellis:1975ap}.
At this low energy, $F_W=7$, $F_\tau = - 4/3$, 
$F_\mu$ decreases smoothly from $-1.41$ to $-1.83$ for $\sqrt{s}=100-200~{\rm MeV}$,
while ${\rm Re}\, F_e$ and ${\rm Im}\, F_e \sim - 10^{-3}$.
For the hadronic contribution, 
Ref.~\cite{Ellis:1975ap} has suggested a value of $F_{\rm had}=20/3$ which we will adopt.
After summing the contributions,
for simplicity we will simply use a constant $F=10$ to calculate the energy loss rate.
In the large $m_r$ limiting case, our estimate is
\begin{equation}
   Q_{\gamma \gamma \rightarrow \alpr \alpr} = \left( \frac{1}{2!} \right)^2\,
   \frac{1819.8}{5 \sqrt{2} \pi}
   \left(\frac{\alpha}{2\pi} \right)^2\, G_F\, |F|^2\, \frac{g^2 \left<\varphi \right>^2}
   {m^4_\varphi\, m^4_r}\, T^{13}\, ,
\end{equation}
where the symmetry factor $(1/2!)^2$ is included for identical particles in the initial
and in the final state.
The estimated energy loss rate translates into an emissivity of
\begin{equation}
   \epsilon_{\gamma \gamma \rightarrow \alpr \alpr} = 
   \frac{Q_{\gamma \gamma \rightarrow \alpr \alpr}}{\rho} =
   \frac{3.94 \cdot 10^{-28}~{\rm GeV}}
   {(\rho / 3 \cdot 10^{14}~{\rm g}/{\rm cm}^3 )}\,\, 
   g^2\, \left(\frac{m_r}{500~{\rm MeV}} \right)^{-4}\, 
   \left(\frac{T}{30~{\rm MeV}} \right)^{13}\, .   
\end{equation}
This is just about the same size as that due to the electron-positron annihilation process, 
thus we conclude that
the photon scattering process does not affect the neutrino cooling time either.

Now we turn to evaluate the energy loss rate due to the nuclear bremsstrahlung process
\begin{eqnarray}
\label{eq:Q_formula}
  Q_{n n \rightarrow n n \alpr \alpr} &=& \frac{\mathcal{S}}{2!} 
  \int \frac{d^3 \vec{q_1}}{2 \omega_1\, (2 \pi)^3}
  \frac{d^3 \vec{q_2}}{2 \omega_2\, (2 \pi)^3}\, \int \prod^{4}_{i=1}
  \frac{d^3 \vec{p_i}}{2 E_i\, (2 \pi)^3}\, f_1 f_2 (1-f_3) (1-f_4) \nonumber \\
   && \hspace{-1cm} \times\, \sum_{\rm spins} 
  |\mathcal{M}_{nn \rightarrow nn \alpr \alpr}|^2\, (2 \pi)^4 
  \delta^4 (p_1 + p_2 - p_3 - p_4 - q_1 - q_2)\, (\omega_1 + \omega_2)\, ,
\end{eqnarray}
where $p_{1, 2}$ are the four-momenta of the initial-state nucleons and $p_{3, 4}$ those
of the final-state nucleons.
For $nn$ or $pp$ interactions, the symmetry factor for identical particles 
is $\mathcal{S}= \frac{1}{4}$, whereas for $np$ interactions it is 1.
The amplitude squared $|\mathcal{M}_{nn \rightarrow nn \alpr \alpr}|^2$ is summed over
initial and final nucleon spins but without being averaged.
In the non-relativistic limit, the occupation numbers are given by the Maxwell-Boltzmann
distribution $f ({\vec{p}}) = (n_B / 2)\, (2 \pi / m_N T)^{3/2}\, e^{- \vec{p}^2 / 2 m_N T}$.
It is normalised such that $\int 2 f (\vec{p}) d^3 \vec{p} / (2 \pi)^3 = n_B$ gives the nucleon
(baryon) density, where the factor 2 accounts for the two spin states.

To calculate the scattering amplitude,
first we need to obtain the effective coupling of the Goldstone bosons to the nucleons.
Recall the Higgs-fermion coupling in SM is
\begin{equation}
  \mathcal{L}_{h f f} = \frac{-1}{\left<\varphi \right>} \sum_f \bar{f} f h\, ,
\end{equation}
with $f= q, Q, l$ denoting the light quarks, heavy quarks, and leptons.
To calculate the Higgs coupling to the nucleon $N = p, n$, 
we need the following matrix element
\begin{equation}
  \left< N | \sum_q m_q\, \bar{q} q + \sum_Q m_Q\, \bar{Q} Q | N \right>\, .
\end{equation}
We will follow the Shifman-Vainshtein-Zakharov (SVZ) approach~\cite{Shifman:1978bx,Shifman:1978by}.
Phenomenologically, at zero momentum transfer one has
$\langle N | Q_\mu^\mu | N \rangle = m_N \bar{\psi}_N \psi_N$, where
\begin{equation}
\label{Qmumu}
   Q_\mu^\mu = \frac{\beta (\alpha_s)}{4 \alpha_s} G^a_{\mu \nu} G^{a\, \mu \nu} + 
   \sum_q m_q\, \bar{q} q + \sum_Q m_Q\, \bar{Q} Q\, ,
\end{equation}
is the energy-momentum tensor,
with $\alpha_s$ the strong coupling constant.
Using the SVZ heavy quark expansion
\begin{equation}
   \sum_Q m_Q\, \bar{Q} Q \rightarrow - \frac{2}{3} \frac{\alpha_s}{8 \pi} n_h\, 
   G^a_{\mu \nu} G^{a\, \mu \nu}\, ,
\end{equation}
and substitute this term into Eq.~(\ref{Qmumu}), in the $m_q \rightarrow 0$ limit we obtain
\begin{equation}
   \langle N | \mathcal{L}_{h f f} | N \rangle \rightarrow - \frac{2}{27} n_h\, 
   m_N \bar{\psi}_N \psi_N\, 
   \cdot \frac{h}{\left<\varphi \right>}\, ,
\end{equation}
where $n_h$ denotes the number of heavy quarks.
The effective Lagrangian for the interaction of Weinberg's Goldstone bosons with the nucleons  
is then
\begin{equation}
   \mathcal{L}_{\rm eff}=\frac{2}{27}\, n_h\, g\, \frac{m_N}{m^2_r m^2_\varphi}\,
   \partial_\mu \alpr\, \partial^\mu \alpr\, \bar{\psi}_N \psi_N
   \equiv \frac{g_N\, m_N}{m^2_r m^2_\varphi}\,
   \partial_\mu \alpr\, \partial^\mu \alpr\, \bar{\psi}_N \psi_N\, .
\end{equation}

\begin{figure}[t!]
\centering
\includegraphics[scale=0.8]{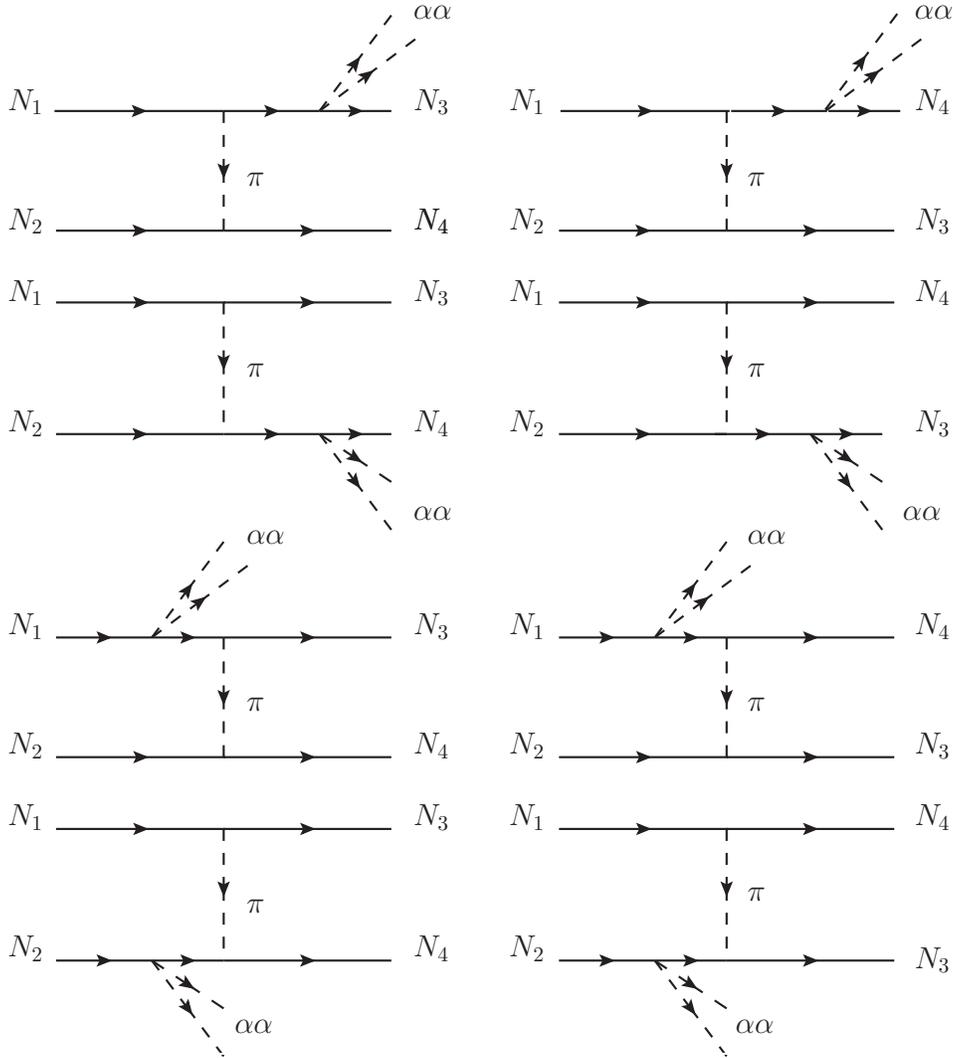}
\caption{Direct (left) and exchange (right) diagrams for the nuclear bremsstrahlung processes 
$N N \rightarrow N N \alpr \alpr$.}
\label{fig:GB_NBr}
\end{figure}

Armed with this knowledge, we follow the prescription given in Ref.~\cite{Brinkmann:1988vi} to calculate the 
amplitude for the nuclear bremsstrahlung process.
In the one-pion exchange (OPE) approximation, there are four direct and four exchange diagrams,
corresponding to the Goldstone boson pairs being emitted by any one of the nucleons
(see Fig.~\ref{fig:GB_NBr}).
In total there are 64 diagrams to calculate, which can be grouped into 8 categories.
Denote the 4-momenta of the exchanged
pions be $k_a \equiv p_2 - p_4$ (in the direct diagrams) and 
$l_a \equiv p_2 - p_3$ (in the exchange diagrams), respectively.
In young supernova cores, $k_a^2 \simeq - |\vec{k}|^2$, $l_a^2 \simeq - |\vec{l}|^2$, 
and $|\vec{k}|^2, |\vec{l}|^2 \sim 3 m_N T$.
Again we work in the large $m_r$ limiting case, using only the leading
term in Eq.~(\ref{mr_large}).
Roughly speaking, the Goldstone bosons are emitted with an average energy of 
several times of $T$, thus for $m_r$ around $400-500~{\rm MeV}$ this approximation is acceptable.
Summing all diagrams from the 8 categories and expanding in powers of $(T / m_N)$,
we find the amplitude squared for the nuclear bremsstrahlung process 
$nn \rightarrow nn \alpr \alpr$ to be 
\begin{eqnarray}
\label{eq:M_nb}
   \hspace{-1cm} \sum_{\rm spins} |\mathcal{M}_{nn \rightarrow nn \alpr \alpr}|^2 &\approx&
   \left(\frac{g_N\, m_N}{m^2_r m^2_\varphi}\right)^2\, \left(\frac{2 m_N f}{m_\pi}\right)^4\, 
   (q_1   \cdot q_2)^2\, \frac{(2 q^2)^2}{(2 p \cdot q)^4}\, m^2_N \nonumber \\
   && \hspace{-2.5cm} \times\, 256 \left\{ \frac{|\vk|^4}{(|\vk|^2 + m^2_\pi)^2} + 
   \frac{|\vl|^4}{(|\vl|^2 + m^2_\pi)^2} +
   \frac{|\vk|^2 |\vl|^2 - 2 |\vk \cdot \vl|^2}{(|\vk|^2 + m^2_\pi) 
   (|\vl|^2 + m^2_\pi)} + ... \right\}\, \cdot (2!)^2\, ,
\end{eqnarray}
with $q = q_1 + q_2$.
Here, $\alpha_\pi \equiv (2 m_N f/m_\pi)^2 /(4 \pi) \approx 15$ with $f \approx 1$ being the 
pion-nucleon ``fine-structure" constant.
The $(2!)^2$ factor accounts for the Wick contraction of the two Goldstone bosons in the
final state.
Considering only the leading terms in the $(T/m_N)$ expansion of the amplitude squared and
neglecting the pion mass $m_\pi$ in the curly brackets,
the phase space integral in Eq.~(\ref{eq:Q_formula}) can be performed analytically
as for the axion or neutrino emission cases~\cite{Raffelt:1993ix}.
We estimate the energy loss rate due to $nn \rightarrow nn \alpr \alpr$ in the non-degenerate 
(ND) case to be
\begin{equation}
   \hspace{-1cm} Q^{\rm ND}_{nn \rightarrow nn \alpr \alpr} 
   \simeq \frac{1056\, \sqrt{\pi}}{(2 \pi)^6} \left(3 - \frac{2 \beta}{3}\right)\, 
   n^2_B\, \left(\frac{g_N\, m_N}{m^2_r m^2_\varphi}\right)^2\,
   \left(\frac{2 m_N f}{m_\pi}\right)^4\, \frac{T^{9.5}}{m^{4.5}_N}\; .
\end{equation}
The $\beta$ term arises from the averaging of the $(\vkdotvl)$ term over the nucleon scattering
angle and we find that $\beta = 2.0938$.
In the large $m_r$ limiting case,
the very strong temperature dependence arises from the presence of the 
$(q_1 \cdot q_2)^2 / m^4_r$ term in
the amplitude squared because of the $\partial_\mu \alpr\, \partial^\mu \alpr \bar{f} f$ type 
coupling~\cite{Freitas} in Eq.~(\ref{ealpha}).
In comparison, in the ND limit the temperature dependence of the energy loss rate is 
$T^{3.5}$ and $T^{5.5}$ for the axion and the neutrino emission cases, 
respectively~\cite{Brinkmann:1988vi,Raffelt:1993ix}.
We compare the emissivity due to the Goldstone bosons 
\begin{equation}
   \epsilon^{\rm ND}_{nn\rightarrow nn \alpr \alpr} = 
   \frac{Q^{\rm ND}_{nn \rightarrow nn \alpr \alpr}}{\rho} 
   \simeq \frac{6.65 \cdot 10^{-22}~{\rm GeV}}
   {\left(\rho / 3 \cdot 10^{14}~{\rm g}/{\rm cm}^3 \right)}\,
   g^2_N\, \left(\frac{m_r}{500~{\rm MeV}} \right)^{-4} 
   \left(\frac{T}{30~{\rm MeV}} \right)^{9.5}\, ,
\end{equation}
with the emissivity bound in Eq.~(\ref{emissivity_bound}), which should be applied at 
$\rho = 3 \cdot 10^{14}~{\rm g}/{\rm cm}^3$ and $T=30$ MeV~\cite{Raffelt:2006cw}.
We obtain a constraint of
\begin{equation}
   g^2_N \left(\frac{m_r}{500~{\rm MeV}} \right)^{-4} \lesssim 1.1 \cdot 10^{-5}\, ,    
\end{equation}
on the coupling of Weinberg's Goldstone bosons to nucleons through the Higgs.
This implies for the coupling constant (cf. Eq.~(\ref{Lagrangian})) 
to the Higgs that
\begin{equation}
   |g| \lesssim 0.011\, \left(\frac{m_r}{500~{\rm MeV}} \right)^2\, ,
\label{keyresult}
\end{equation}
from the relation $g_N = (2/27)\, n_h\, g$, with the number of heavy quark
flavours $n_h=4$. 
One sees that the supernova bound is competitive and complementary to the collider bound 
$g \lesssim 0.018\, (0.011)$, which is insensitive to the $m_r$ value.
We have checked the pion mass effects on the energy loss rate by
keeping the $m^2_\pi$ in the denominators in Eq.~(\ref{eq:M_nb}) and performing the phase space
integrals using the Monte Carlo routine VEGAS~\cite{Lepage:1977sw}.
We find that the reduction is $12\%$ at $T=30$ MeV and only $5\%$ at $T=80$ MeV, milder than
that in the axion emission case. 
It remains to estimate the emissivity for more general cases, 
i.e. for smaller $m_r$ values, and including
the higher-order terms in the $(T/m_N)$ expansion of the amplitude squared
(Eq.~(\ref{eq:M_nb})), to find the modifications of this bound.

In conclusion, we have determined the allowed range for the coupling constant $g$ in dependence of 
$m_r$, the mass of the radial field $r (x)$ in Weinberg's extended Higgs model, 
in which new Goldstone bosons 
may be masquerading as fractional cosmic neutrinos. 
In the large $m_r$ limiting case,
we have estimated the energy loss rates in post-collapse supernova cores due to Goldstone boson emission in different channels including the $e^+ e^-$ annihilation, photon scattering and nuclear
bremsstrahlung processes.
We present our main result in Eq.~(\ref{keyresult}), obtained by  
confronting our estimate for the nuclear bremsstrahlung processes with the well established emissivity bound from the Supernova 1987A observations and simulations. 
For $m_r$ large enough compared with the temperature in the post-collapse 
supernova core, our bound is very competitive to that derived from collider experiments.
Technical details, investigation of more general cases, as well as other astrophysical constraints will be presented in a following work~\cite{longpaper}.

This work was supported in part by the National Science Council, Taiwan,
ROC under the NSC Grant Nos. 101-2112-M-001-010-MY3  (KWN, HT), 
101-2112-M-001-005-MY3 (TCY),
and in part by U.S. Department of Energy under the Grant DE-FG02-12ER41811 (WYK).

\end{document}